\documentclass[twocolumn]{aastex631}
\usepackage{amsmath}     
\usepackage{wasysym}           
\usepackage{graphicx}
\usepackage{amssymb}
\usepackage[outdir=./]{epstopdf}
\usepackage{mathrsfs}

\usepackage{anyfontsize}
\usepackage{natbib}

\usepackage{natbib}
\usepackage{color}
\usepackage{lipsum}
\usepackage{diagbox}

\shortauthors{Cufari, Coughlin, \& Nixon}
\begin{document}

\title{Using the Hills Mechanism to Generate Repeating Partial Tidal Disruption Events and ASASSN-14ko}

\author[0000-0001-8429-754X]{M. Cufari}
\affiliation{Department of Physics, Syracuse University, Syracuse, NY 13210, USA}
\email{mcufari@syr.edu}
\author[0000-0003-3765-6401]{Eric R.~Coughlin}
\affiliation{Department of Physics, Syracuse University, Syracuse, NY 13210, USA}
\email{ecoughli@syr.edu}
\author[0000-0002-2137-4146]{C.~J.~Nixon}
\affiliation{Department of Physics and Astronomy, University of Leicester, Leicester, LE1 7RH, UK}

\begin{abstract}
Periodic nuclear transients have been detected with increasing frequency, with one such system -- ASASSN-14ko -- exhibiting highly regular outbursts on a timescale of $114 \pm 1$ days. It has been postulated that the outbursts from this source are generated by the repeated partial disruption of a star, but how the star was placed onto such a tightly bound orbit about the supermassive black hole remains unclear. Here we use analytic arguments and three-body integrations to demonstrate that the Hills mechanism, where a binary system is destroyed by the tides of the black hole, can lead to the capture of a star on a $\sim 114$ day orbit and with a pericenter distance that is comparable to the tidal radius of one of the stars within the binary. Thus, Hills capture can produce stars on tightly bound orbits that undergo repeated partial disruption, leading to a viable mechanism for generating not only the outbursts detected from ASASSN-14ko, but for periodic nuclear transients in general. We also show that the rate of change of the period of the captured star due to gravitational-wave emission is likely too small to produce the observed value for ASASSN-14ko, indicating that in this system there must be additional effects that contribute to the decay of the orbit. In general, however, gravitational-wave emission can be important for limiting the lifetimes of these systems, and could produce observable period decay rates in future events. 
\end{abstract}


\section{Introduction} \label{sec:introduction}
Repeating optical and X-ray flares have recently been detected in the nuclei of galaxies, e.g., \citet{miniutti19, song20, chakraborty21,arcodia21, payne21, arcodia21, payne21b}, with ASASSN-14ko in particular displaying outbursts at an extremely regular period of $P = 114$ days (\citealt{payne21}; see \citet{shappee14, holoien17, kochanek17} for specifics related to the ASAS-SN survey). The physical origin of these flares (in general) remains an open question, but if the flaring activity is related to accretion onto the supermassive black hole (SMBH) in the galactic nucleus, the high degree of regularity in the flaring rate of ASASSN-14ko (specifically) is strongly indicative of a binary origin. \citet{payne21} suggested that this binary companion (to the SMBH) is a star that is partially disrupted upon each pericenter passage, and the accretion of the tidally liberated debris produces the observed flares (see, e.g., \citealt{lacy82, rees88, gezari21} and references therein for overviews of the tidal disruption process and observational implications).

However, if the star was fed to the SMBH through traditional, two-body scattering and thus approached the SMBH on a $\sim$ parabolic orbit (e.g., \citealt{frank76, lightman77, cohn78, stone16}), the question arises as to how the partially disrupted star dissipated enough kinetic energy to be bound to the black hole with a period as short as $\sim 100$ days. If part of the star survived the tidal encounter, then the maximum energy able to be imparted to it (i.e., without completely destroying it) is $E \simeq GM_{\star}^2/R_{\star}$ (which agrees with the recent simulation results of \citealt{nixon21}; see their Equations 5 -- 8), and hence the orbital period of the star after its interaction with the SMBH would be (assuming that all of the energy imparted to the star comes at the expense of the center of mass motion)

\begin{equation}
    P_{\bullet} \simeq P_{\rm ff}\left(\frac{M_{\bullet}}{M_{\star}}\right), \label{Pbullet1}
\end{equation}
where

\begin{equation}
    P_{\rm ff} = \frac{\pi R_{\star}^{3/2}}{\sqrt{2GM_{\star}}} \label{Pff}
\end{equation}
is, to within a factor of 2, the freefall time from the stellar surface. For solar values, $P_{\rm ff} \simeq 1$ hr, and so for a SMBH mass of $10^{7}M_{\odot}$ we have $P_{\bullet} \simeq 10^{3}$ yr. Thus, it seems necessary for some mechanism to \emph{place} the partially tidally disrupted star on its $\sim 114$ day orbit about the SMBH, and particularly to reduce the period derived above by a factor of $\sim 1000$.

Here we propose Hills capture as that mechanism, and specifically that the star being repeatedly partially disrupted was originally part of a binary system and captured -- and placed on a relatively tight orbit -- by the SMBH; this mechanism, as we show below, yields a timescale that is reduced by the factor of $\sim 1000$ necessary to reproduce the period observed in ASASSN-14ko. The possibility that the captured star resulting from the Hills mechanism could itself be repeatedly partially disrupted by the SMBH, but in a different region of parameter space than that considered here, was also analyzed by \citet{antonini11}. Section \ref{sec:analytic} presents analytic calculations that substantiate this possibility and show that we can reproduce the observed properties of ASASSN-14ko with this mechanism, Section \ref{sec:three} gives the results of three-body integrations to substantiate the estimates in Section \ref{sec:analytic}, and we discuss the implications of our findings and conclude in Section \ref{sec:conclusion}.

\section{Analytic Estimates}
\label{sec:analytic}
A binary of semimajor axis $a_{\star}$ and primary mass $M_{\star}$ that nears a SMBH of mass $M_{\bullet}$ will be destroyed if it comes within a distance of approximately \citep{hills88}

\begin{equation}
    r_{\rm t} = a_{\star}\left(\frac{M_{\bullet}}{M_{\star}}\right)^{1/3} \label{rt}
\end{equation}
of the SMBH. Within this tidal radius the tidal field of the black hole overwhelms the self-gravity of the binary. The difference in energy across the binary (the tidal potential) implies that one binary member is ejected, while the other is captured on an orbit with semimajor axis

\begin{equation}
    a_{\bullet} \simeq \frac{a_{\star}}{2}\left(\frac{M_{\bullet}}{M_{\star}}\right)^{2/3} \label{abullet}
\end{equation}
and period 

\begin{equation}
    P_{\bullet} \simeq P_{\star}\left(\frac{M_{\bullet}}{M_{\star}}\right)^{1/2}, \label{Tbullet}
\end{equation}
with $P_{\star} = \pi a_{\star}^{3/2}/\sqrt{2GM_{\star}}$ (compare to Equations \ref{Pbullet1} and \ref{Pff} above for the case of a single star that \emph{survives} the tidal encounter). In the nucleus of a galaxy, the high velocity dispersion $\sigma \simeq 200$ km s$^{-1}$ implies that binaries must have 

\begin{equation}
    a_{\star} \lesssim \frac{GM_{\star}}{\sigma^2} \simeq 0.02\left(\frac{M_{\star}}{M_{\odot}}\right)\left(\frac{\sigma}{200\textrm{ km s}^{-1}}\right)^{-2}\textrm{ AU}.
\end{equation}
If we take a black hole mass of $M_{\bullet} = 10^{7} M_{\odot}$, a primary mass of $M_{\star} = 1M_{\odot}$, and a binary separation of $a_{\star} = 0.005$ AU, then Equation \eqref{Tbullet} gives

\begin{equation}
    P_{\bullet} \simeq 144\textrm{ days} \label{Tbullet2}
\end{equation}
for the orbital period of the captured star, which is $\sim$ the flaring period of ASASSN-14ko.

From Equations \eqref{rt} and \eqref{abullet}, the eccentricity of the orbit of the captured star is

\begin{equation}
    e_{\bullet} \simeq 1-\frac{2}{\beta}\left(\frac{M_{\bullet}}{M_{\star}}\right)^{-1/3} \simeq 0.991, \label{ebullet}
\end{equation}
where we defined $\beta \equiv r_{\rm t}/r_{\rm p}$ with $r_{\rm p}$ the pericenter distance of the binary's orbit about the SMBH (and we require $\beta \gtrsim 1$ for the successful disruption of the binary; see Section \ref{sec:three} below). From \citet{peters63}, the time derivative of the orbital period of the captured star due to gravitational wave emission is

\begin{multline}
    \dot{P}_{\bullet} 
    = -\frac{192\pi G^3}{5 c^5}\sqrt{\frac{a_{\bullet}}{GM_{\bullet}}}\frac{M_{\bullet}^2M_{\star}}{a_{\bullet}^3\left(1-e_{\bullet}^2\right)^{7/2}}\left(1+\frac{73}{24}e_{\bullet}^2+\frac{37}{96}e_{\bullet}^{4}\right)\\
    \simeq -\frac{85\pi}{8}\frac{a_{\star}^{3/2}}{\sqrt{2GM_{\star}}}\frac{G^3M_{\star}^2M_\bullet}{c^5a_{\star}^4}\beta^{7/2} \simeq -1.3\times10^{-6} \label{gw}.
\end{multline}
Here we used Equations \eqref{abullet} and \eqref{ebullet} to remove the dependence on $a_{\bullet}$ and $e_{\bullet}$, we set $M_{\bullet} = 10^7M_{\star}$, $M_{\star} = 1M_{\odot}$, $\beta = 1$, and $a_{\star} = 0.005$ AU in the last equality, and we approximated $1-e_{\bullet}^2\simeq 2\left(1-e_{\bullet}\right)$ and set $1+73/24e_{\bullet}^2+37/96e_{\bullet}^4 \simeq 425/96$. This is about a factor of 1000 smaller than the value inferred from the observations of ASASSN-14ko, and hence an additional mechanism may be necessary to reduce the orbit in that system (e.g., interaction with the accretion flow; \citealt{syer99}); the disagreement between the expected gravitational-wave decay rate and that observed in ASASSN-14ko was also noted by \citet{payne21}. We note that Equation \eqref{rt} with $a_{\star} = 0.005$ AU and $M_{\bullet}/M_{\star} = 10^{7}$ is

\begin{equation}
    r_{\rm t} \simeq 5.5\textrm{ R}_{\rm g},
\end{equation}
where $R_{\rm g} = GM_{\bullet}/c^2$. Thus, the disruption process of the binary takes place very near the horizon of the black hole, and as a consequence we require that $\beta \lesssim 2$. For $\beta \gtrsim 2$, the SMBH directly captures the binary. The small separation of the binary -- which is necessary to survive the extreme environment in the nucleus of the galaxy -- also implies that 

\begin{equation}
    a_{\star} = 0.005\textrm{ AU} \simeq 1.1 R_{\odot},
\end{equation}
meaning that (1) if the disrupted star is $\sim$ solar with a radius $\lesssim R_{\odot}$, the other object must be compact and likely must be a compact object to avoid being in a common-envelope phase, and (2) the tidal disruption radius of the binary is $\sim$ the tidal disruption radius of the solar-like star. The latter implies that the tidal disruption of the binary will likely produce at least a partial disruption of the star itself\footnote{Recent simulations have shown that $\beta \gg 1$ can also result in the reformation of a stellar core following the initial stellar disruption \citep{nixon22}, which could produce recurrent flares in more extreme versions of this type of Hills capture, i.e., with smaller black hole masses (to avoid direct capture) and larger $\beta$.} and generate the flares. 

The analysis here demonstrates that the Hills capture of a star can generate a repeating, partial TDE with similar overall characteristics to ASASSN-14ko. Moreover, the relativistic nature of the pericenter of the star implies that the debris stream undergoes a relativistic apsidal precession angle that is $\gtrsim 180^{\circ}$ upon returning, and hence we would expect rapid circularization of the disrupted material. In the next section we produce statistics from the Hills disruption of binaries to substantiate the estimates made here.

\section{Three-body Integrations}
\label{sec:three}
Here we analyze the results of Newtonian three-body interactions between a tight binary and a SMBH. The binary is injected at an initial distance of $50\,r_{\rm t}$ such that the center of mass is on a parabolic orbit about the SMBH with pericenter distance $r_{\rm p} = r_{\rm t}/\beta$, where $r_{\rm t}$ is given by Equation \eqref{rt}, and $\beta = 0.5$, 1, 2, and 4. The plane of the binary is randomly oriented with respect to the orbital plane of the binary center of mass. The binary is initially circular and has a mass ratio of one. Successful captures are recorded when the distance between the two stars exceeds $50\,a_{\star}$. All length scales can be measured relative to $a_{\star}$, all timescales relative to $a_{\star}^{3/2}/\sqrt{GM_{\bullet}}$, and all masses relative $M_{\bullet}$, but we restrict the analysis to $a_{\star} = 0.005$ AU and $M_{\bullet} = 10^{7}M_{\odot}$ for concreteness and to compare to the previous section. 

\begin{figure*}
\includegraphics[width=0.495\textwidth]{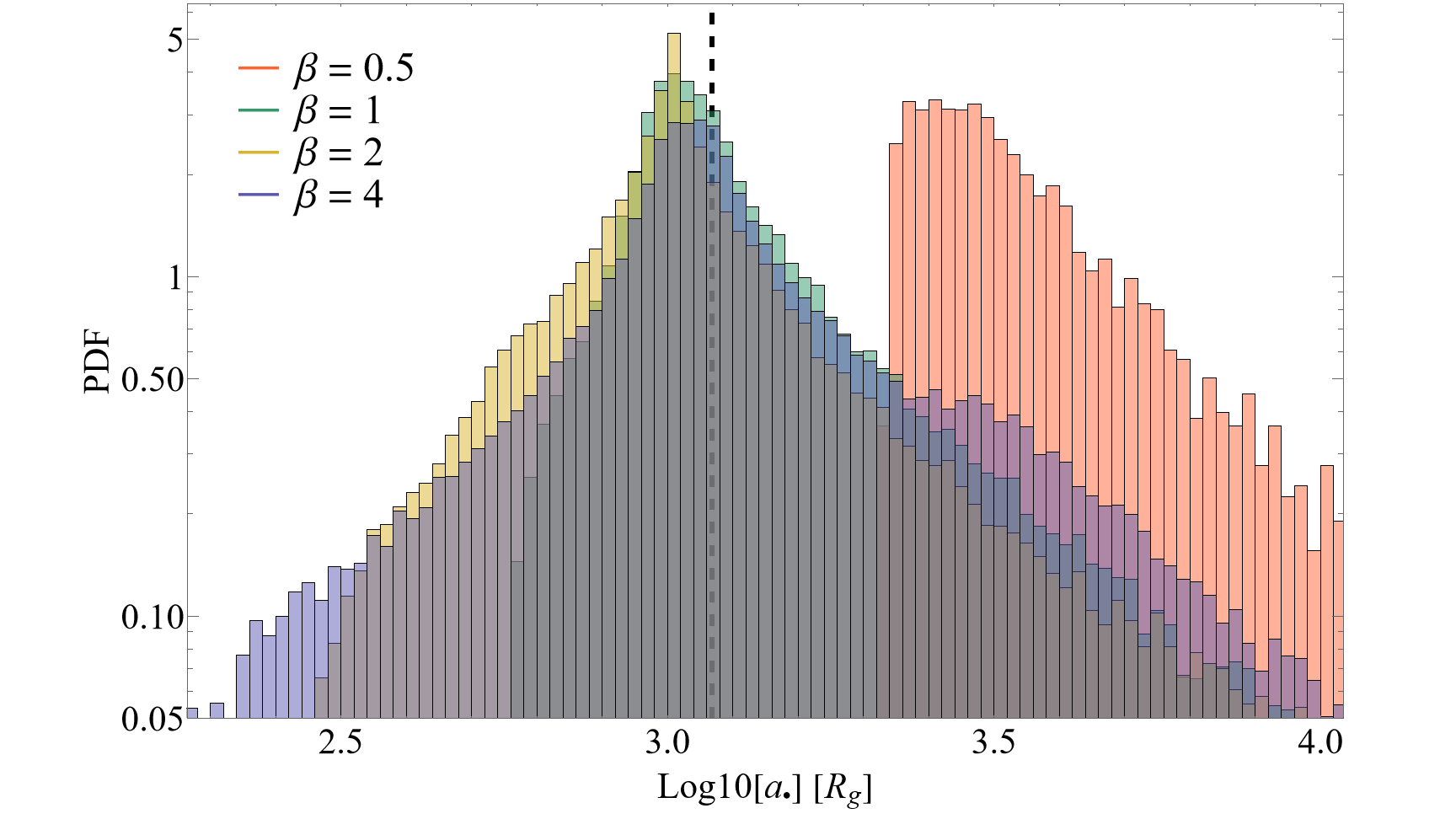}
\includegraphics[width=0.495\textwidth]{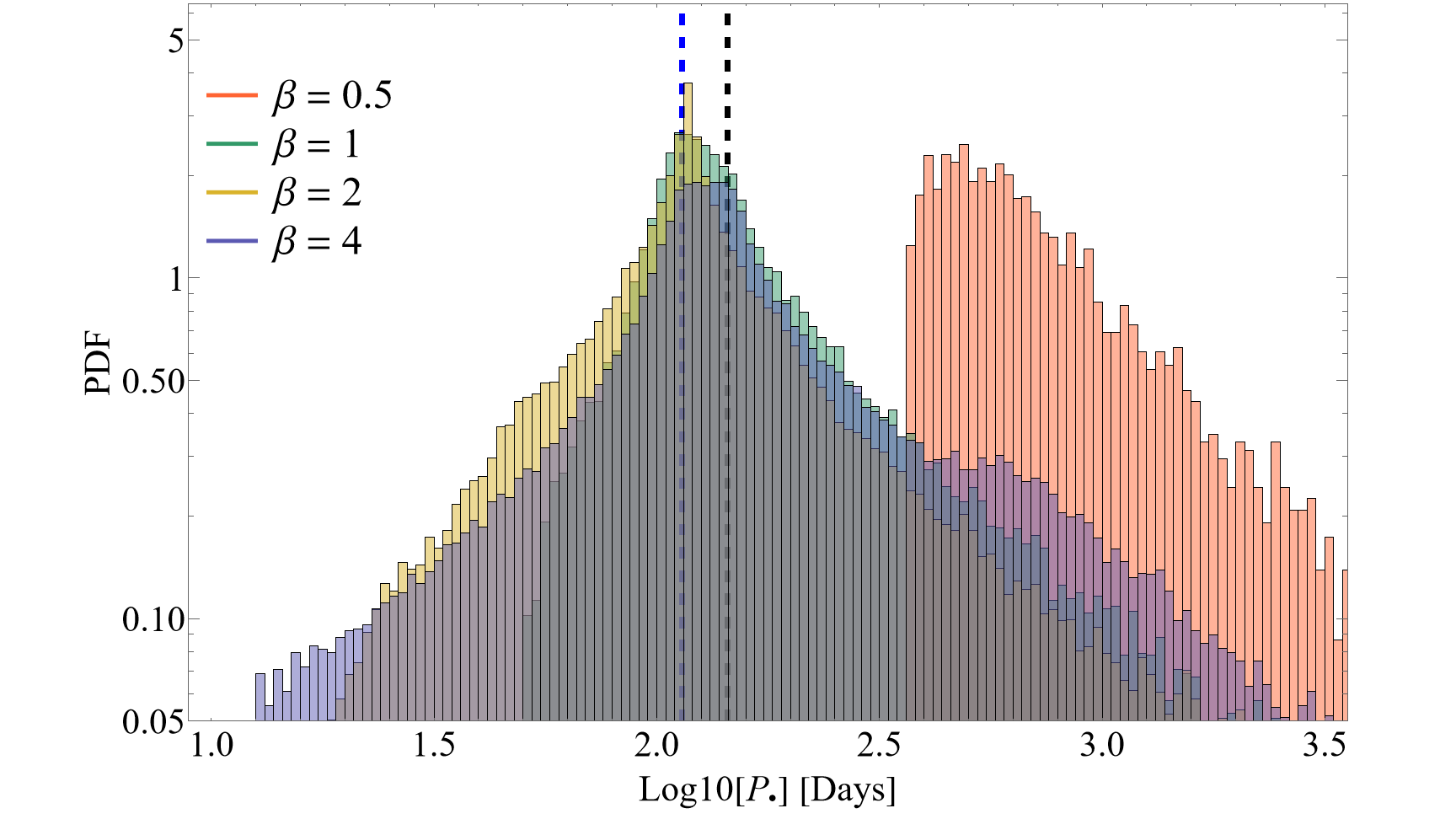}
\includegraphics[width=0.495\textwidth]{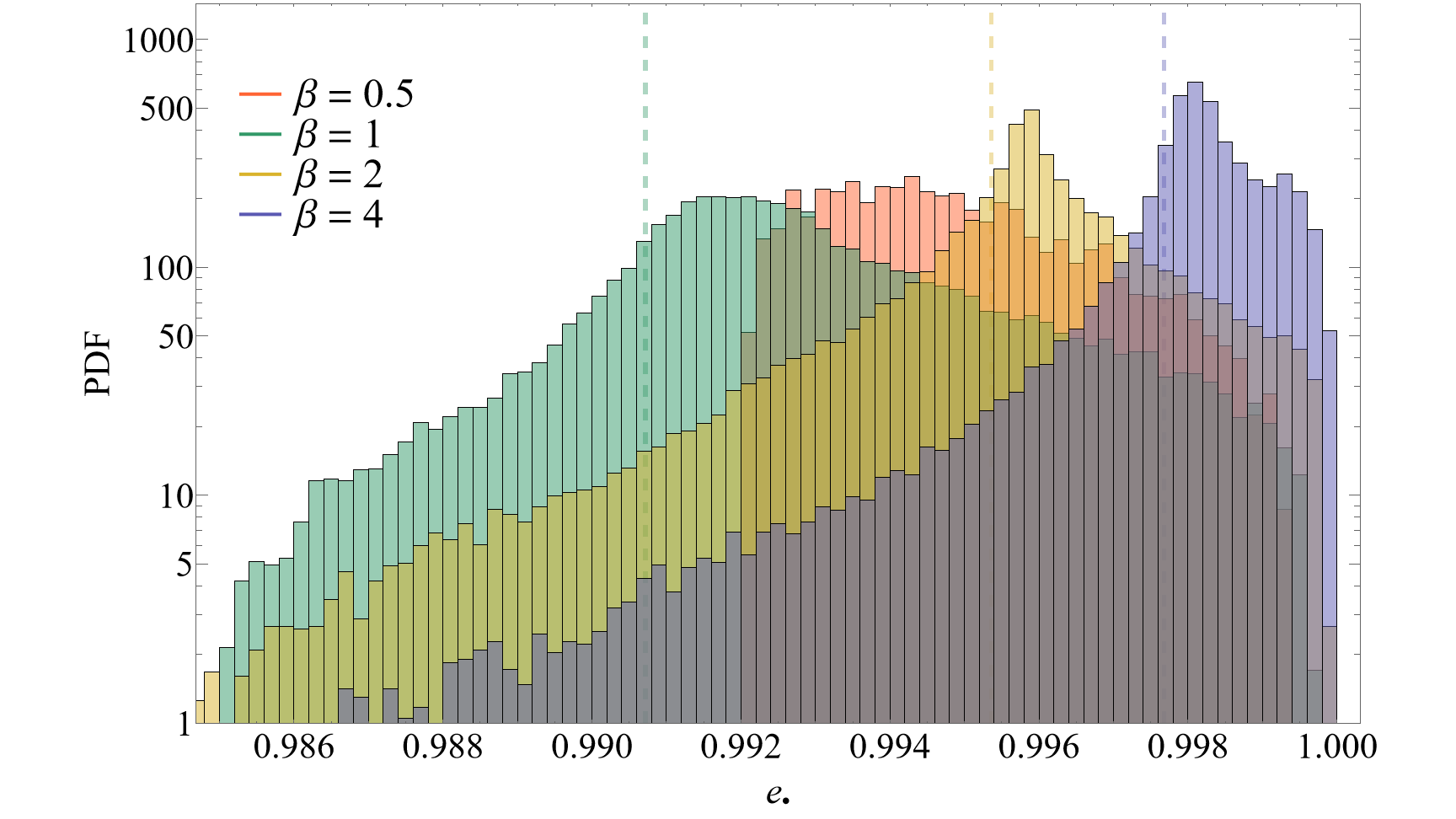}
\includegraphics[width=0.495\textwidth]{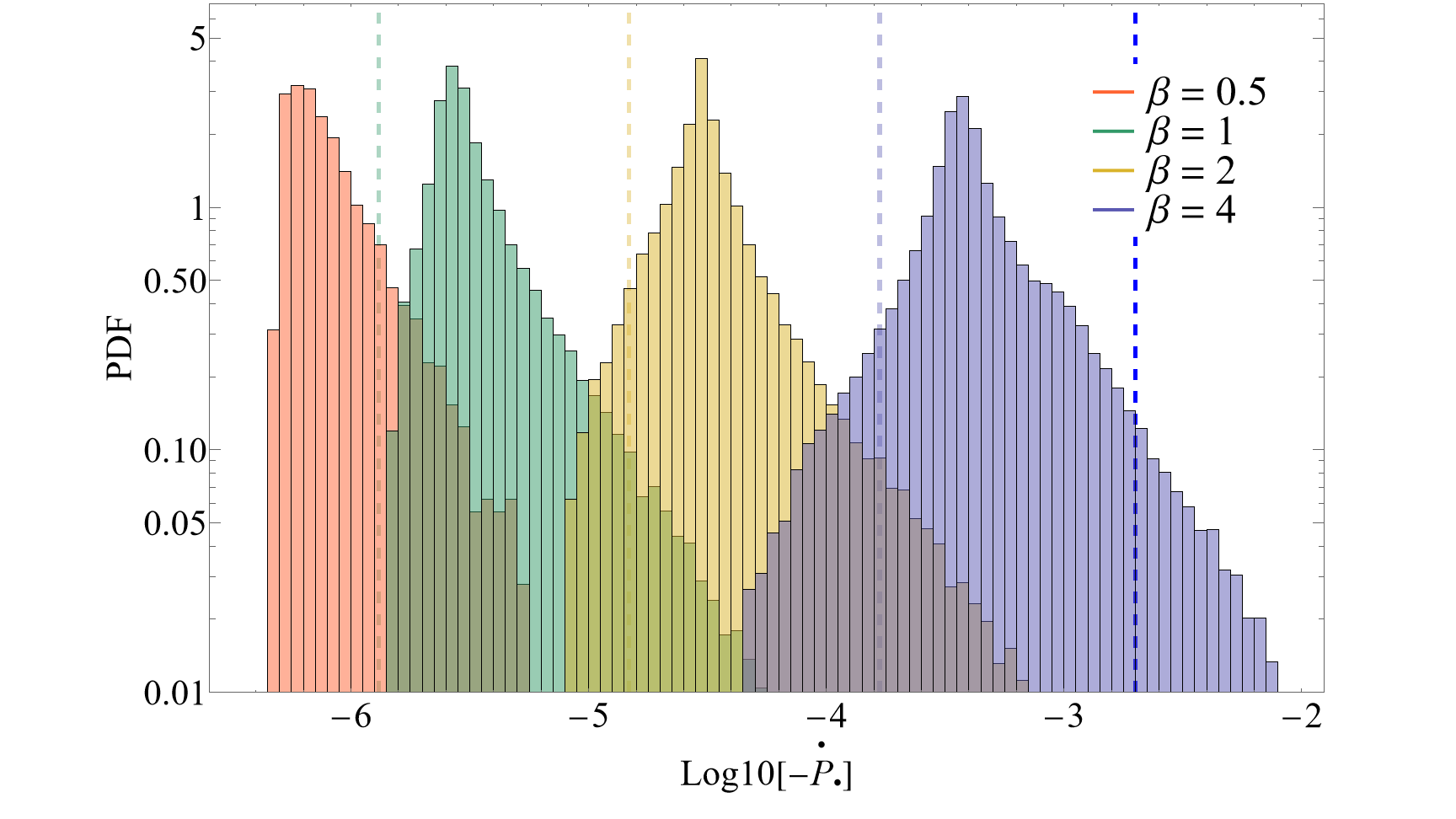}
\caption{The probability distribution function (PDF) of the semimajor axis of the captured star, $a_\bullet$ (top-left), the orbital period of the captured star, $P_\bullet$ (top-right), the eccentricity of the captured-star orbit, $e_\bullet$ (bottom-left), and the period decay rate due to the emission of gravitational waves, $\dot{P}_\bullet$ (bottom-right), for an initial binary semimajor axis of $a_{\star} = 0.005$ AU and SMBH mass of $M_{\bullet} = 10^{7}M_{\odot}$. Vertical, dashed lines on these plots give the estimates that result from the calculations in Section \ref{sec:analytic}; the predicted eccentricity for $\beta = 0.5$ is $e_{\bullet} \simeq 0.981$, and was excluded from the bottom-left panel to retain the clarity of the figure. We similarly excluded the vertical, dashed line for $\beta = 0.5$ ($\dot{P}_\bullet = -1.15\times 10^{-7}$) in the bottom right panel. The different colors correspond to the $\beta$ shown in the legend, where $\beta$ is the ratio of the tidal radius of the binary to the point of closest approach of the binary. The vertical blue line in the top-right corresponds to the period of ASASSN-14ko, being 114 days, and the vertical, blue line in the bottom-right panel is the observed $\dot{P}_{\bullet}$ of ASASSN-14ko, being $\dot{P} = -0.002$.}
\label{fig:results}
\end{figure*}

The three-body integrations were performed with a package integrator in Mathematica; the integrator preserved the orbital energy of test (i.e., only two-body, Newtonian) binaries to within a factor of $\Delta E/E \lesssim 10^{-7}$, where $\Delta E/E$ is the fractional change in the energy of the binary per orbit, over the time taken for the binary to reach the pericenter of its orbit. In previous works \citep{darbha18, darbha19}, similar three-body integrations were also compared to the $N$-body code {\sc rebound} with a 15th-order accurate integration scheme \citep{rein12, rein15}, and we found excellent agreement between the results. Our statistics (Table \ref{tab:1}) and probability distributions (Figure \ref{fig:results}) are also in agreement with those reported in previous studies, e.g., \citet{hills88, brown18}.

\begin{table*}[]
    \centering
    \begin{tabular}{|c c c c c c|}
    \hline \hline 
      $\beta$ & Capture \% & $\mu_{a_{\bullet}}$ ($\sigma_{a_{\bullet}}$) [$R_{\rm g}$] & $\mu_{P_{\bullet}}$ ($\sigma_{P_{\bullet}}$) [Days] & $\mu_{e_{\bullet}}$ ($\sigma_{e_{\bullet}}$) & $\mu_{\dot{P}_{\bullet}}$ ($\sigma_{\dot{P}_{\bullet}}$) \\
       \hline 
       0.5 & 2.88 & 2601 (1390) & 481.8 (378.4) & 0.9942 (0.004424)  & $-5.96\times 10^{-7}~ (3.20\times10^{-7})$\\
       \hline 
       1.0 & 55.8 & 1005 (390.9) & 122.9 (72.79) & 0.9918 (0.003516) & $-2.63\times 10^{-6}~ (1.01\times 10^{-6})$ \\
       \hline 
       2.0 & 71.5 & 1014 (255.4) & 115.2 (48.92) & 0.9958 (0.000988) & $-3.01 \times 10^{-5}~ (7.95\times10^{-6})$ \\
       \hline 
       4.0 & 81.2 & 1047 (471.2) & 119.8 (86.80) & 0.9981 (0.000893) & $-3.72 \times 10^{-4}~ (1.64\times 10^{-4})$ \\
       \hline
    \end{tabular}
    \caption{Capture percentage and the peak locations, $\mu$, with corresponding full width at half maximum, $\sigma$, for each distribution shown in Figure \ref{fig:results}.}
    \label{tab:1}
\end{table*}

Figure \ref{fig:results} shows the probability distributions of semimajor axis of the captured star (CS) in units of $R_{\rm g}$, the orbital timescale of the CS in days, the eccentricity of the CS orbit, and the change in orbital timescale due to the emission of gravitational waves, $\dot{P}_\bullet$. Table \ref{tab:1} gives the peaks ($\mu$) and full-width at half-maxima ($\sigma$) of the PDFs shown in this figure; the first column in this table gives the capture probability (e.g., 55.8\% of the encounters between the binary and the SMBH lead to the disruption of the binary for $\beta = 1$). These results show that the approximations in Section \ref{sec:analytic} are highly accurate and that the orbital timescale and semimajor axis of the CS are effectively independent of $\beta$ for $\beta > 1$, in agreement with Equations \eqref{abullet} and \eqref{Tbullet}.

The disruptions with $\beta = 0.5$, on the other hand, do not yield the same level of qualitative agreement with the predictions in Section \ref{sec:analytic}. The origin of this discrepancy is that the analytic arguments are only expected to be valid for $\beta \gtrsim 1$, i.e., when the tidal force exceeds the self-gravity of the binary. The disruptions for $\beta = 0.5$ (which only constitute $\lesssim 3\%$ of the encounters, as shown in Table \ref{tab:1}) therefore arise from fortuitous arrangements of the binary near pericenter, and are only produced when the binary orbital plane is nearly aligned with the orbital plane of the binary center of mass, because the alignment between the two increases the effective tidal radius of the system (see \citealt{brown18}; see also \citealt{golightly19} for the analogous conclusion in the case of a single star disrupted by an SMBH).

\section{Discussion and Conclusions}
\label{sec:conclusion}
The periodic nuclear transient ASASSN-14ko was recently argued by \citet{payne21} to be a repeating, partial TDE, where the star is on an orbit with a period of $\sim 114$ days and is partially disrupted -- producing an accretion flare -- once per orbit. This scenario requires a mechanism to bind the star to the SMBH and place it on its $\sim 114$ day orbit. Here we demonstrated, through the estimates in Section \ref{sec:analytic} and the results of three-body integrations in Section \ref{sec:three}, that this mechanism can be the Hills capture of a star in a binary system. Specifically, a star in a binary with semi-major axis $a_{\star} \sim 0.005$ AU (necessarily of this order because of the high velocity dispersion in the nucleus of the galaxy) disrupted by a $10^{7}M_{\odot}$ SMBH yields a captured star with a most likely period of $\sim 120$ days. Assuming one of the members of the binary has a radius $\lesssim 1 R_{\odot}$ implies that the pericenter distance of the binary is $\sim$ the tidal radius of the star, and hence this set of parameters also results in the partial disruption of one of the stars. 

The Hills mechanism provides a general means of placing stars on tightly bound orbits within the partial disruption radius of the captured star. This process could, for example, be applicable to systems that display quasi-periodic eruptions on much shorter timescales, e.g., those observed in the nucleus of GSN 069 \citep{miniutti19}, provided that the radius of the star is correspondingly much smaller. \citet{king20} recently suggested that GSN 069 could be powered by a white dwarf being partially disrupted by the SMBH in that system. Applying the analysis in Section 2 to this system suggests that the Hills mechanism may be a possible route to capturing a white dwarf into the required orbit (in contrast, \citealt{king20} suggested that the white dwarf is a relic of a past, partial disruption of a red giant).

However, a caveat with this possible origin of the short-timescale, quasi-periodic eruptions is that the SMBH mass must be small to avoid swallowing the white dwarf. For a stellar radius of 0.02 $R_{\odot}$ and a stellar mass of $0.21 M_{\odot}$ \citep{king20}, the direct capture radius ($4 R_{\rm g}$) coincides with the tidal radius when $M_{\bullet} \simeq 2.5\times 10^{5} M_{\odot}$, which is slightly smaller than the value inferred for GSN 069. However, when the tidal disruption radius is relativistic, partial disruptions can occur at larger radii than would be predicted in the Newtonian limit (e.g., \citealt{kesden12, gafton15}). Additionally, the star will be rapidly rotating in a prograde sense as a consequence of the tidal torque of the SMBH, which further facilitates the disruption of the star at larger radii than would be expected in the irrotational case \citep{golightly19}. Thus, partial disruptions could be more readily achievable owing to strong-gravity (relativistic) effects and the rotation of the star. 

The statistical analysis in Section \ref{sec:three} shows that there is a fair amount of scatter around the estimates for the orbital properties (e.g., the orbital period of the captured star) that were obtained in Section \ref{sec:analytic}. However, the PDF declines dramatically for periods less than $\sim P_{\bullet}/2$ when $\beta = 1$, as shown in Figure \ref{fig:results}. The timescales are also sensitive to the value of the binary separation, and scale as $a_{\star}^{3/2}$. Thus, increasing the semimajor axis by a factor of 2 (and keeping the other parameters unchanged) implies that it is statistically nearly impossible to achieve an orbital period as short as $114$ days for $\beta = 1$. Furthermore, decreasing the semimajor axis by a factor of 2 places the tidal radius within the direct capture radius of the SMBH. This model is therefore highly constraining as concerns the orbital parameters of the initial binary. 

Because the binary is very tight, the companion star must be sufficiently compact so as to avoid being in a common envelope phase. If the companion star is a white dwarf, this negates the possibility of the companion also being disrupted (or partially disrupted) by the SMBH, which would be possible if both stars were solar-like and as described in \citet{mandel15}. The fallback time associated with the debris should also be associated with the canonical fallback time given by Equation \eqref{Tbullet2} (because the stellar radius is $\sim$ the binary semimajor axis); this is a factor of a few longer than the observed duration of the flares from ASASSN-14ko, being $\sim 30-40$ days. However, simulations of partial tidal disruption events find that the flare duration (e.g., the time to peak) can be significantly shorter than that of full events (e.g., \citealt{guillochon2013, nixon21}; see Figures 1 and 2 of \citealt{nixon21}). The fallback (accretion) rate is also reduced by a factor of 10-100 compared to the value one would expect in a full disruption (again, \citealt{guillochon2013, nixon21}; see Figures 1 and 2 of \citealt{nixon21}), which is in good agreement with the peak luminosity observed in ASASSN-14ko of $\sim few\times 0.01 \, L_{\rm Edd}$ \citep{payne21}.

We assumed that the disrupted star is a main-sequence star with $R_{\star} \lesssim R_{\odot}$ because the tight semimajor axis of the binary (necessitated by the high velocity dispersion in the galactic nucleus) would put a more massive star in a common-envelope phase with its companion. If the disrupted star were a giant, its outer, tenuous envelope could be more easily stripped by the SMBH \citep{macleod13}, but the arguments in Section \ref{sec:introduction} still imply that the tidal dissipation timescale would be too long to produce the tight orbit of the captured star. Low-mass stars are also significantly more abundant by number, though the stellar initial mass function may be somewhat top-heavy in galactic nuclei (\citealt{kroupa01}; but see \citealt{lockmann10}).

The reduction in the orbital period of the captured star owing to gravitational-wave emission for our fiducial set of parameters is $\dot{P}_{\bullet} \simeq - few \times 10^{-6}$, as shown in the bottom-right panel of Figure \ref{fig:results} for $\beta = 1$. Increasing the value of the impact parameter (i.e., reducing the pericenter distance of the original binary) significantly reduces the gravitational-wave inspiral time, and with $\beta = 2$ the value is a factor of $\sim 10$ larger than that inferred for ASASSN-14ko, being $\dot{P} \simeq -0.002$. This disagreement could imply that the period decay in ASASSN-14ko is related to a distinct physical origin (i.e., not associated with gravitational-wave emission alone), such as the interaction between the star and the AGN disc \citep{syer99}. Nevertheless, the gravitational-wave inspiral time is short for our fiducial parameters ($\lesssim 10^6$ years), and will limit the lifetimes of these Hills-captured, repeating-partial TDEs for large black hole masses and small binary semimajor axes. 

The arguments in the Introduction (see the text around Equation \ref{Pbullet1}) suggest that single stars that are partially disrupted by SMBHs can achieve a minimum orbital timescale of $P_{\bullet} \simeq P_{\star}\left(M_{\bullet}/M_{\star}\right)$, where $P_{\star}$ is $\sim$ the dynamical time of the star. For typical systems with solar-like parameters and $P_{\star} \lesssim 1$ hr $\sim 10^{-4}$ yr, this timescale is $\sim 100$ -- $10^{4}$ yr for SMBH masses in the range $10^{6}$ -- $10^{8} M_{\odot}$. Thus, it seems unlikely that nuclear transients that repeat on humanly accessible timescales and that are powered in this way -- by a star being repeatedly partially disrupted -- could be populated by single star-SMBH interactions, and instead must be generated by another means (such as the Hills mechanism described here).

For the initial study carried out here, we kept the properties of the injected binary at a mass ratio of 1 and an eccentricity of 0. In future work we plan to perform a more exhaustive investigation of the influence of these additional parameters on the properties of the captured star, as well as the influence of relativistic effects on the encounter, to understand the plausibility of this mechanism for providing the origin for flaring events in galactic nuclei. Specifically, while it seems likely that general relativity will not substantially modify the orbit of the captured star in a secular way (the dominant effect is still periapsis advance or, if the SMBH possesses substantial rotation, nodal precession, because the gravitational-wave decay timescale is still $\gg$ the orbital period for typical parameters; see Section \ref{sec:analytic}), it may be that the capture itself is more likely for modest $\beta$ from the increased strength of the relativistic tidal field \citep{kesden12}. Additionally, the mean binding energy of the captured star could be reduced further by general relativistic effects, making short-period orbits more likely. In either case, we anticipate that general relativity will only strengthen the viability of this mechanism for producing ASASSN-14ko-like transients.

\section*{}
 We thank the anonymous referee for useful and constructive comments and suggestions. M.C.~acknowledges this research was supported in part through computational resources provided by Syracuse University and through funding provided by the Syracuse University Office for Undergraduate Research (SOURCE). E.R.C.~and M.C.~acknowledge support from the National Science Foundation through grant AST-2006684 and a National Science Foundation Research Experiences for Undergraduates supplement. C.J.N acknowledges funding from the European Union’s Horizon 2020 research and innovation program under the Marie Skłodowska-Curie grant agreement No 823823 (Dustbusters RISE project).

\bibliographystyle{aasjournal}

\end{document}